\documentclass[10pt,conference]{IEEEtran}
\ifCLASSOPTIONcompsoc
  \usepackage[nocompress]{cite}
\else
  \usepackage{cite}
\fi
\usepackage{prettyref}
\usepackage[boxed,linesnumbered,ruled]{algorithm2e}
\usepackage{algorithmicx}
\usepackage{amsmath}
\usepackage{multirow}
\usepackage{diagbox}
\usepackage{subcaption}
\SetCommentSty{textnormal}
\usepackage{hyperref}
\usepackage[pdftex]{graphicx}
\usepackage{xcolor}
\newrefformat{eqn}{Equation~\ref{#1}}
\newrefformat{fig}{Figure~\ref{#1}}
\newrefformat{tab}{Table~\ref{#1}}
\newrefformat{sec}{Section~\ref{#1}}
\newrefformat{line}{Line~\ref{#1}}
\newrefformat{ssec}{Section~\ref{#1}}
\newrefformat{lst}{Listing~\ref{#1}}
\newrefformat{alg}{Algorithm~\ref{#1}}

\ifCLASSINFOpdf
\else
\fi
\begin{document}
%
\title{ZSMILES: an approach for efficient SMILES storage for random access in Virtual Screening}


\author{\IEEEauthorblockN{Gianmarco Accordi\IEEEauthorrefmark{1},
Davide Gadioli\IEEEauthorrefmark{1}, Giorgio Seguini\IEEEauthorrefmark{1}, 
Andrea R. Beccari\IEEEauthorrefmark{2}, and Gianluca Palermo\IEEEauthorrefmark{1}}
\vspace{0.2cm}
\IEEEauthorblockA{\IEEEauthorrefmark{1}Dipartimento di Elettronica Informatica e Bioingegneria - Politecnico di Milano, Milano, Italy\\
\IEEEauthorblockA{\IEEEauthorrefmark{2}EXSCALATE - Dompé Farmaceutici SpA, Naples, Italy\\
Email: \IEEEauthorrefmark{1}\{firstname.lastname\}@polimi.it \IEEEauthorrefmark{2}andrea.beccari@dompe.com}
}
}

\maketitle

\begin{abstract}
Virtual screening is a technique used in drug discovery to select the most promising molecules to test in a lab.
To perform virtual screening, we need a large set of molecules as input, and storing these molecules can become an issue. 
In fact, extreme-scale high-throughput virtual screening applications require a big dataset of input molecules and produce an even bigger dataset as output.
These molecules' databases occupy tens of TB of storage space, and domain experts frequently sample a small portion of this data.
In this context, SMILES is a popular data format for storing large sets of molecules since it requires significantly less space to represent molecules than other formats (e.g., MOL2, SDF).

This paper proposes an efficient dictionary-based approach to compress SMILES-based datasets. This approach takes advantage of domain knowledge to provide a readable output with separable SMILES, enabling random access.
We examine the benefits of storing these datasets using ZSMILES to reduce the cold storage footprint in HPC systems.
The main contributions concern a custom dictionary-based approach and a data pre-processing step.
  
From experimental results, we can notice how ZSMILES leverage domain knowledge to compress $\times1.13$ more than state of the art in similar scenarios and up to $0.29$ compression ratio.
We tested a CUDA version of ZSMILES targetting NVIDIA's GPUs, showing a potential speedup of $7\times$.
\end{abstract}

\section{Introduction}\label{sec:introduction}
The drug discovery process is a lengthy and costly process for a pharmaceutical company \cite{DRUGDISCOVERYCOST}.
The process produces molecules (called \textit{ligands}), which are more likely to interact with at least one binding site (called \textit{pocket}) on a protein (called the screening's \textit{target})\cite{EXSCALATE}.
Ideally, these interactions inhibit the target's activity, leading to a favorable therapeutic effect.
The process comprises in-silico, in-vitro, and in-vivo stages.

Virtual screening is an in-silico stage, which can improve the success rate \cite{VSDG} of the drug discovery pipeline by selecting chemical compounds with a higher likelihood of interacting with the target protein.
We screen compounds from a large dataset to ensure the most likely interactions \cite{reviewParallelVS,reviewVSHT}.
Virtual screening eliminates the molecules with low binding affinity for in-vitro and later in-vivo testing\cite{chemVS}.

In a virtual screening campaign, evaluating the initial compounds' interaction strength against multiple target proteins is common.
This cartesian product increases the required computation effort.
Moreover, the evaluation of each ligand-protein pair is independent of the others, making the problem embarrassingly parallel.
For these reasons, supercomputers are the ideal target for extreme-scale virtual screening campaigns.

One challenge of virtual screening campaigns is the storage requirement \cite{summitBigrun, exascalateBigrun}, to describe the input chemical library and the output, which usually decorates the input with the strength of their interactions.
For example, the screening data of a virtual screening campaign on CINECA's Marconi100 \cite{exascalateBigrun} was approximately 72 TB.
Not all screening data is accessed on a daily basis, but rather, domain experts sample this chemical space to create a smaller subset.
To mitigate this problem, it is common to encode molecules using the SMILES format, which describes a molecule using a single line of ASCII characters.
Given the rising popularity of large virtual screening campaigns and SMILES format, storing them efficiently is of general interest.

This work presents ZSMILES, a methodology to reduce the SMILES storage footprint of extreme-scale virtual screening applications. 
The proposed approach employs a dictionary-based compression (and decompression).
Leveraging domain knowledge to reduce SMILES's storage footprint is one of the contributions of this paper.
Our approach takes advantage of SMILES format specifications to get better compression. In particular, we have addressed the dictionary generation and data preparation phase.
Moreover, we have other constraints: domain experts have to access these databases easily without, for example, the burden of handling binary characters.
Domain experts must also cut and combine SMILES databases: we defined the shared dictionary to be input-independent and SMILES to remain separable.
Thus, our approach employs a single fixed dictionary to compress any set of SMILES and enhance maintainability and compatibility.
SMILES separability implies each SMILES to be placed on different lines.
ZSMILES's output still provided each SMILES on the same input line number.
In the following with random access, we refer to maintaining input SMILES order in output, thus for SMILES to remain separable.
We implemented ZSMILES in a serial C++ version targeting CPUs.
Large virtual screening experiments are usually accelerated using GPUs.
Thus, we have tested a parallel CUDA version targeting GPUs.

We evaluate the impact of domain-specific optimizations on performance. ZSMILES compression ratios are evaluated with different optimizations enabled and dictionaries trained on different datasets of SMILES. We also analyze ZSMILES's implementations execution times. 

The remainder of this article is structured as follows: initially, we report the background \prettyref{sec:background}.
Subsequently, we provide a state-of-the-art analysis in \prettyref{sec:sota}. 
Next, \prettyref{sec:methodology} outlines the proposed methodology and illustrates the accelerated approach.
Finally, an analysis of the results acquired in section \prettyref{sec:results} is presented, with the conclusions given in \prettyref{sec:conclusions}.
Moreover, abbreviations and technical terms are explicitly defined when first used to facilitate clear communication.

\section{Background}\label{sec:background}
The methodology proposed in this study focuses on SMILES strings, which can describe molecules using a single-line notation that encodes only its topological information using UTF-8 ASCII characters \cite{SMILES}.
Therefore, the proposed approach falls in the category of short-string compression algorithms.
This format provides an excellent balance between human-readable and computationally parsable notations.
The main idea is to start the molecule description from a terminal heavy atom and write all the other heavy atoms attached to the starting one in sequence.
It uses round brackets to handle branches in the molecule structure.
When we encounter a ring, we remove an edge and assign a numerical ID to the attached atoms, and then we resume the encoding procedure.
Therefore, this numerical ID can identify all the molecule rings. 
The SMILES format has additional rules to encode chemical information, such as upper-case chemical symbols for non-aromatic atoms and lower-case chemical symbols for aromatic ones.
From a SMILES string, it is possible to add hydrogens and compute the 3D displacement of their atoms at runtime.

\begin{figure}[t]
  \centering
  \includegraphics[width=\columnwidth]{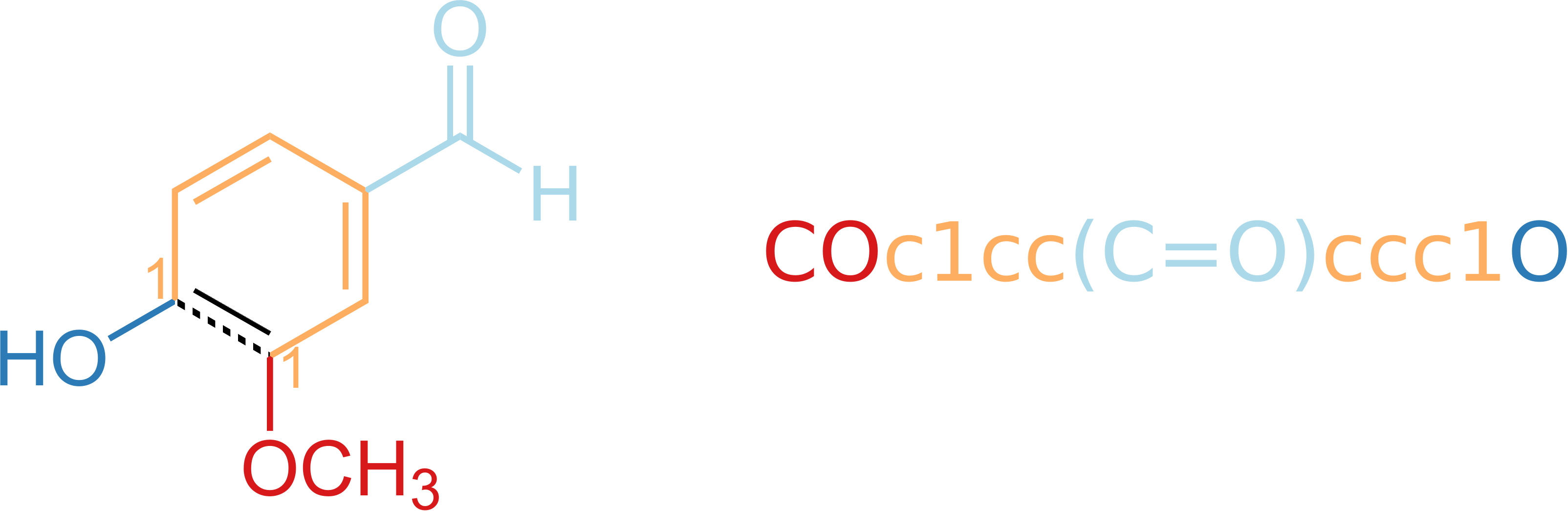}
  \caption{Graphical representation of Vanillin on the left, while on the right, its SMILES representation.}
  \label{fig:vanillin}
\end{figure}
For example, \prettyref{fig:vanillin} shows a graphical representation of the Vanillin, \textit{COc1cc(C=O)ccc1O} in SMILES format.
In this example, the SMILES representation starts from the bottom part of the molecule, i.e., from the lowermost carbon.
Thus, our SMILES notation starts with the $C$ (since this carbon is not aromatic).
Following the molecule structure, the next heavy atom is a non-aromatic oxygen, expanding the SMILES string to $CO$.
The next atom is an aromatic carbon that belongs to a ring.
To represent it, we remove the edge on its left (represented by the dashed lines ), assign ID $1$, and add the carbon to SMILES notation $COc1$.
Then, we resume the encoding structure, including two aromatic carbons, i.e., $COc1cc$.
At this point, we reach a branch that we need to handle using round brackets.
For simplicity, we assume to encode the right branch first, so our SMILES representation becomes $COc1cc(C=0)$ (the $=$ symbol stands for a double covalent bond).
Then, we also encode the left branch, which includes all the remaining heavy atoms, i.e., $COc1cc(C=0)ccc1O$.

Please notice how the SMILES description is not unique; it depends on how we topologically order the molecule structure.
Moreover, the ring enumeration does not have to be unique.
The only requirement is that nested ring descriptions should not have overlapping IDs to prevent ambiguities in their representation.

\section{State of the Art}\label{sec:sota}
Dictionary-based compression algorithms \cite{Gagie2008} work by parsing the input while attempting to match substrings against a predefined set within a dictionary.
When a match occurs, the algorithm substitutes the substring in the input with the correspondent symbol of the dictionary entry: if the latter is shorter than the former, we have reduced the string size.
The approach used for the dictionary generation substantially impacts the compression ratio delivered by the algorithm.
We can also compress data by changing the encoding of the output.
Entropy coders, for example, try to reach the lower bound on the number of bits required to represent the symbols by leveraging the frequency of input patterns. Huffman coding and Lempel-Ziv are examples of entropy coders \cite{4051119}.

SMILES compression can be achieved by using state-of-the-art binary compression tools such as Bzip2\footnote{Website: \url{https://sourceware.org/bzip2/}}, DEFLATE, and LZ77\cite{8117850}, or by relying on tools for short string compression like SMAZ\footnote{Website: \url{https://github.com/antirez/smaz}}, SHOCO\footnote{Website: \url{https://ed-von-schleck.github.io/shoco/}}, and FSST\cite{10.14778/3407790.3407851}.
None of the previously mentioned tools meet our defined requirements: readable output, separable SMILES, and a shared dictionary.
In the experimental results, we compare ZSMILES with Bzip2, as a representative of binary compressors file-based, and with FSST and SHOCO, as representative of small string compression.

Bzip2 compresses files using multiple layers ranging from the Burrows-Wheeler and move-to-front transform \cite{10.1145/382780.382782} to the Huffman coding \cite{4051119}.
Its compression ratio is high compared to other algorithms at the expense of time and computation \cite{8117850}. Bzip2 compression is stateful, so to decompress one part of the file, it also has to decompress the previous part; thus, random access is impossible.
Therefore, to meet our use case requirements, we can use Bzip2 to compress SMILES files line by line, which makes SMILES compression inefficient since the input string needs to be larger to reach a good compression ratio.
Nonetheless, the compressed SMILES with Bzip2 are in binary format, so they are not human-readable.

SHOCO\footnote{Website: \url{https://ed-von-schleck.github.io/shoco/}} is a compression library for short strings.
It provides a way to generate a custom dictionary based on the application domain. 
Since it is an entropy encoder, it provides a non-readable and non-random access output.
FSST instead offers a good compression ratio on small strings within a dataset while allowing random access to the file.
FSST uses a static symbol table defined from a small chunk of data from the input file.
Since the table is static, it is immutable during the compression and decompression.
FSST constructs a symbol table for each input; thus, the dictionary is input-dependent.
In addition to that, FSST's output is non-readable since the dictionary uses non-printable ASCII symbols, which can cause problems with third-party tools.

Another work available in the literature which leverages domain knowledge to achieve better SMILES compression through data preprocessing was presented by Gupta et al. \cite{Scanlon2013}. It targets compression using file-wide binary compression tools such as Bzip2. This approach does not apply to our case because the compressed file cannot be accessed line-by-line, which does not guarantee the possibility of random access.


\section{Methodology}\label{sec:methodology}
This section introduces the proposed ZSMILES approach.
The main idea is to use a dictionary that associates a common string pattern to an ASCII character.
To compress SMILES, we substitute all the dictionary patterns in SMILES with the related characters.
We escape in output any characters in the input SMILES that cannot be encoded using the dictionary.
The first two sections explain two optimizations that hinge on domain knowledge to improve the quality of the dictionary.
Then, \prettyref{sec:dictionary} describes how we generate the dictionary in more detail, and \prettyref{sec:execflow} reports the compression and decompression algorithm.
Ultimately, we detail the CUDA accelerated SMILES compression approach in \prettyref{sec:cuda}.

\subsection{Preprocessing}
\label{sec:preprocessing}
\begin{figure}[t]
  \centering
  \includegraphics[width=\columnwidth]{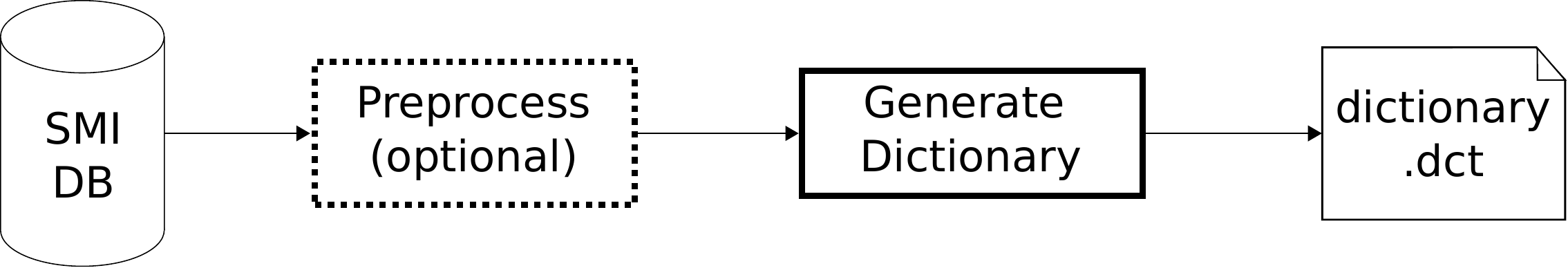}
  \caption{Graphical representation of the SMILES pre-processing step.}
  \label{fig:dictionary}
\end{figure}
SMILES pre-processing aims to increase the probability of finding common patterns. SMILES use numbers (IDs) to identify the opening and closing of rings.
Since ring IDs are not reused, it became more difficult to identify common patterns.
We propose pre-processing the inputs to increase the reuse of ring enumerations and the probability of finding common patterns.
We can reuse ring IDs in 2 ways: innermost or outermost.
If multiple rings are nested within each other, we can give the lower ID to the innermost or the outermost.
We chose the innermost approach because the simplest and most common rings are the inner ones, which have the smaller ID values.

Take as an example the Dibenzoylmethane SMILES representation:

\texttt{C1=CC=C(C=C1)C(=O)CC(=O)C2=CC=CC=C2}

\noindent this SMILES have two rings, \textit{C1=CC=C(C=C1)} and \textit{C2=CC=CC=C2}, which have a similar prefix, but with different IDs.
Thus, compressing both of them would require the dictionary to have two entries representing the same prefix.
Instead, if pre-processed, the SMILES becomes:

\texttt{C0=CC=C(C=C0)C(=O)CC(=O)C0=CC=CC=C0}

\noindent which now enables the compression of the SMILES by leveraging the occurrences of \textit{C0=CC=C}.
It is worth noting that the pre-processed SMILES remain valid.

\subsection{Dictionary Pre-population}
\label{sec:perpop}
If the input SMILES has a pattern not included in the generated dictionary, we must escape all its characters, doubling the required size.
Usually, known approaches mitigate this issue by using compression algorithms that build a dictionary tailored to the target input file.
However, we would like to use a shared dictionary in our use case, increasing the chances of missing a pattern.
For this reason, we can pre-populate the dictionary with the printable ASCII characters used by the SMILES format, thus avoiding escaping.
For example, we use the character \texttt{@} for chiral specification, the character \texttt{/} for stereoisomers specification, or the character \texttt{\#} for triple bonds.
This conservative choice reduces the number of characters representing patterns to the extended ASCII characters.
However, when the input SMILES is compliant with the format, we have the guarantee that the compressed file does not require more storage than the input one.
%
In the experimental results, we measured the compression ratio increment of this optimization.

\subsection{Dictionary Generation}
\label{sec:dictionary}
\prettyref{fig:dictionary} shows how ZSMILES generates the dictionary $D$, from an input training set of SMILES.
The training set of SMILES is parsed to find a group of $T$ recurrent substrings of the inputs that provide the highest coverage of the inputs.
The coverage measures how much of the input is covered by the available substrings.
This problem can, however, be seen as a knapsack problem \cite{Kellerer2004}, whose complexity is NP-complete. 

\begin{algorithm}[t]
  \SetKwInOut{Parameter}{Parameters}
  \KwIn{SMILES $\gets$ training set}
  \KwOut{D $\gets$ substring dictionary}
  \Parameter{$L_{min} \gets$ Minimum substring length\\$L_{max} \gets$ Maximum substring length \\ $C \gets$ Initial dictionary values \\ $T \gets$ Dictionary size }
  $rank \gets \{\}$\; \label{line:dictGenOne}
  $D \gets C$\; \label{line:dictGenTwo}
  \ForEach{$line$  $\in$ SMILES }{ \label{line:dictGenThree}
    \ForEach{$s$  $\in$ $line$ : $L_{min} \leq |s| \leq L_{max}$}{
    $rank[s] = \begin{cases}
        rank[s]+1$ if $ s \in rank,\\
        1 $ otherwise$
    \end{cases}$
    }
  } \label{line:dictGenSeven}
  \ForEach{$t$ $\in \{0, 1, \cdots, T\}$}{ \label{line:dictGenEight}
    new\_word$ = \operatorname*{argmax}_{rank(s)} s \in rank$\;
    $D \gets D \cup \{ $ new\_word $ \}$\;
    $rank \gets rank\ \setminus $ new\_word\;
    \ForEach{$s \in rank$}{ 
      $update\_rank(s)$\;
    }
  } \label{line:dictGenFifteen}
  \caption{Dictionary generation algorithm.}
  \label{alg:dictGen}
\end{algorithm}
\prettyref{alg:dictGen} is the pseudocode of how dictionaries are generated.
The input is the training set of SMILES, while the output is the dictionary.
The input is parsed by searching for unique substrings with length in the interval $[L_{min}, L_{max}]$.
We used a maximum substring length of $L_{max}$ because longer substrings require a lot of time to generate the dictionary.
$L_{min}$ was set to $2$ to speed up the dictionary generation based on the previous detailed dictionary pre-population.
\prettyref{line:dictGenOne} initializes $rank$, which contains the rank of all substrings found, while \prettyref{line:dictGenTwo} initializes $D$, the compression dictionary.
We define the $rank$ of a pattern $p$ at step $t$ as the product between its occurrences in the input, times its normalized length $occ_{pattern} \times norm_{p,t}$, as in \prettyref{eqn:rank}:
\begin{equation}
  \label{eqn:rank}
  rank_{p,t} = occ_{p} \times ( l_{p} - overlap_{p,t} )
\end{equation} 
Where $norm_{p,t}$ of pattern $p$ at step $t$ is defined as the difference between the length $l_{p}$ of $p$ and the overlap with patterns selected in the previous iteration of the loop in \prettyref{line:dictGenEight}.
From \prettyref{line:dictGenThree} to \prettyref{line:dictGenSeven}, count the occurrences of all substrings in the input.
From \prettyref{line:dictGenEight} to \prettyref{line:dictGenFifteen}, we populate  $D$ (size $T$), and at each iteration, we get the pattern that has the higher rank. 
Then, all other ranks are updated based on the pattern selected at that step.

\subsection{ZSMILES Execution Flow}
\label{sec:execflow}
\begin{figure*}[t]
  \centering
  \includegraphics[width=0.8\textwidth]{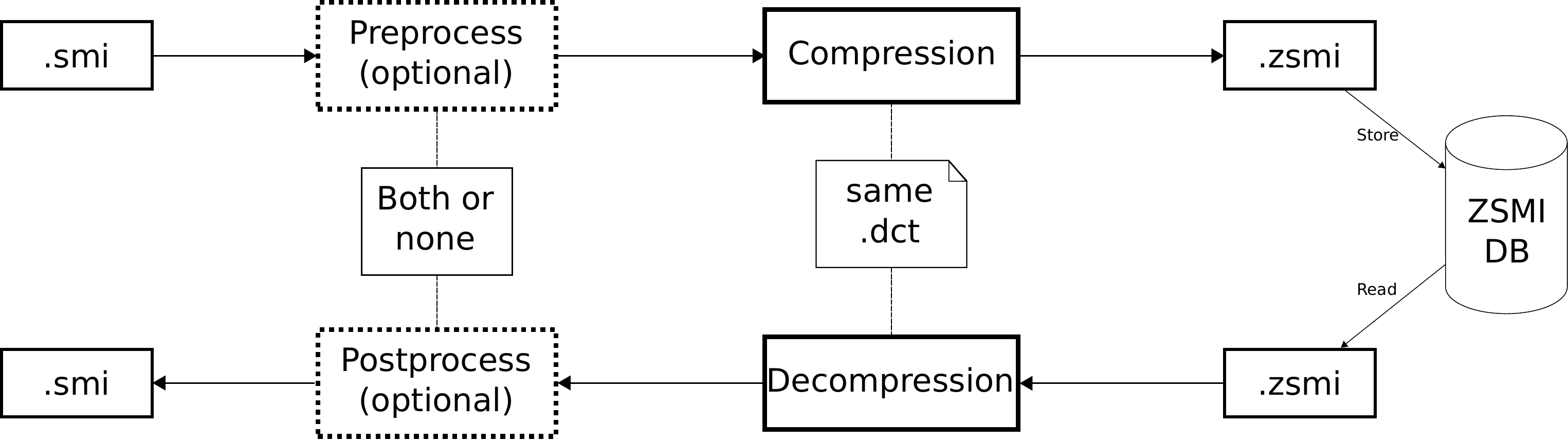}
  \caption{Graphical representation of the compression and decompression process in ZSMILES.}
  \label{fig:alg}
\end{figure*}
\prettyref{fig:alg} exemplify ZSMILES compression and decompression process.
The flow in the upper part of the diagram illustrates the compression process: ZSMILES optionally pre-process the input before compressing and storing it.
The lower part of the diagram instead reports the decompression process of ZSMILES, which works backward: from storage, compressed SMILES are decompressed and optionally post-processed.
The dictionary is soft-coded in the ZSMILES executable, so we cannot change it once ZSMILES is compiled.

\subsubsection{Compression Algorithm}
\label{sec:comprAlg}  
We formulate the compression algorithm as an optimization problem.
The input is a single line from a SMILES file.
The output is the compressed SMILES.
The problem is identifying the sequence of substrings in $D$ that gives the best compression ratio of the input.
Dictionary $D$ is represented by a trie \cite{10.1145/367390.367400} to do pattern matching on the input symbols.
We use the Dijkstra algorithm to get the shortest path from the first char in SMILES up to the last one.
To apply Dijkstra, we need to construct a graph $G$ of the SMILES: each node $n$ is an input character, and an edge $(n_1, n_2)$ represents a substring in $D$ which begins with character $n_1$ and ends with character $n_2$.
The algorithm scans the graph $G$ to match patterns from the trie of $D$.
At each iteration $i$, the Dijkstra algorithm computes the best way to compress the input from character $n_i$ up to the input end, based on a cost function.
Each match (represented by a graph edge) costs one if it is from a symbol in $D$.
Otherwise, the cost is two if the symbol has to be escaped with a trailing \textit{space} because there was no match in $D$.
The last iteration calculates the shortest path to compress the input SMILES and traverses it from the beginning while printing the symbol of the corresponding pattern match on the output.

\subsubsection{Decompression}
\label{sec:decomprAlg}
The decompression algorithm is instead straightforward.
During decompression, we use the dictionary $D$ as a lookup table.
For each symbol in each compressed SMILES, we perform a lookup in $D$ and print out its expansion.
If, instead, the value is a \textit{space}, due to the escaping, we go to the next symbol in the input and print this symbol directly.

\subsection{CUDA Implementation}
\label{sec:cuda}
To reduce ZSMILES compression (and decompression) overhead, we have developed a CUDA implementation, which allows us to exploit the computational power offered by NVIDIA GPUs.
Given an input set of SMILES, the compression or decompression is split among groups of CUDA \textit{threads}, called \textit{blocks}, which build up a CUDA \textit{grid}.

For compression, each block compresses a SMILES.
Each block's thread looks at different input SMILES's characters: for each dictionary element, the thread checks if the correspondent substrings can be matched in the input, starting from that character.
In this way, the block constructs a graph representation of the SMILES: a substring match creates an edge, which connects the substring's first character node with the node of the last character.
We apply weights on the edges using a cost function in the same way as described in \prettyref{sec:comprAlg}.
Once a block has a graph representation of the input SMILES, it scans the graph backward by applying Dijkstra.
The shortest path identified by Dijkstra gives the best compression ratio of the input SMILES.

For decompression, each block decompresses a SMILES.
Each block's thread performs a lookup into the dictionary using an input SMILES character.
Therefore, each thread knows the decompressed string's dimension for each input character.
Finally, block threads share how many characters they must write in output in order to know where to start the writing.

We have set blocks to have the same dimension as a CUDA warp.
A warp is a set of 32 CUDA threads with specific scheduling properties.
In our case, we rely on warps' synchronization and shuffle CUDA operations for performance reasons.

\section{Results}\label{sec:results}
In this section, we discuss and report the experimental results of ZSMILES.
We provide details of the setup and data used in \prettyref{sec:expSet}.
Then, we analyze and discuss ZSMILES based on the compression ratios (\prettyref{sec:compRatio}) and the performance (\prettyref{sec:perf}).

\subsection{Experimental Setup}
\label{sec:expSet}
Our experiment used a machine equipped with an AMD EPYC 7282 16-core processor, 64 GB of RAM, and two NVIDIA A100 graphics cards.

Since the dictionary generation has to be input-independent, we need a heterogeneous set of databases.
Thus, we used three datasets that we consider representative: two of them, GDB-17\cite{10.1021/ci300415d} and MEDIATE database\cite{Vistoli2023} are public, while one of them is a set of SMILES from a real case virtual screening execution \cite{EXSCALATE}, that we call EXSCALATE.
GDB-17 contains 166 billion small organic molecules.
MEDIATE is a dataset of ligands from commercial compounds to natural products.
We also used a MIXED dataset to construct the dictionary, combining the first one million ligands from each dataset.
Dataset and additional material will be available on a public repository on GitHub \footnote{Website: \url{https://github.com/elvispolimi/zmsiles}}.

\subsection{Compression Ratio}
\label{sec:compRatio}
In this section, we report the results of different experiments on the ZSMILES compression ratio.

\textbf{Dictionary Optimizations.}
The first experiment evaluates the ZSMILES compression ratio with different dictionary generation mechanisms.
In particular, we evaluate the impact of what has been proposed in \prettyref{sec:preprocessing} and in \prettyref{sec:perpop}.
To train the dictionaries used in \prettyref{tab:zsmilesRatio}, we used a sample of random $50000$ SMILES from the mixed dataset, the same one used to test the compression ratio in the experiment.
The experiment collects ZSMILES compression ratios with all combinations of optimizations.
We compare the compression ratio of ZSMILES when the dictionary is pre-populated with all the printable ASCII characters, with the characters of the SMILES alphabet, or with no characters at all.
We expect a better compression ratio when the dictionary is initialized with a subset of common characters in SMILES. 
 
\begin{table}[b]
  \centering
  \caption{Compression ratios of ZSMILES using different dictionaries.}
  \label{tab:zsmilesRatio}
  \begin{tabular}{|c|c|c|}
    \hline
    \textbf{Pre-processing} & \textbf{Pre-population} & \textbf{Compression Ratio} \\
    \hline
    Yes                         & Printable     & 0.32                       \\
    \hline
    No                          & Printable     & 0.35                       \\
    \hline   
    Yes                         & SMILES alphabet        & 0.29                       \\
    \hline
    No                          & SMILES alphabet        & 0.32                       \\
    \hline
    Yes                         & None                & 0.33                       \\
    \hline
    No                          & None                & 0.35                       \\
    \hline
  \end{tabular}
\end{table}
\prettyref{tab:zsmilesRatio} reports the experiment's results.
The first column indicates whether the pre-processing has been done on the input data. The second column reports the set of ASCII characters used to pre-populate the dictionary.
From \prettyref{tab:zsmilesRatio}, we can see how, in all cases, the reuse of ring ID has improved the compression ratio, and we can also see that pre-populating the dictionary with the SMILES alphabet provides a better compression ratio, up to $0.29$.

In the remaining part of this work, we apply pre-processing before generating all the dictionaries that are initialized with the SMILES alphabet.
These are the proposed optimization techniques.

\textbf{Cross-dictionary.}
The second experiment we conducted aims to find the best dataset to use for dictionary training.
The experiment analyses the tradeoffs of using the same shared dictionary for any input set of SMILES.
We have evaluated ZSMILES compression ratios with dictionaries trained on each available dataset against all others.
We expect the compression to be worse when the training set consists of similar SMILES and better when we try to compress a dataset using a dictionary trained on the same one.

\begin{table}[b]
  \addtolength{\tabcolsep}{-2pt}
  \centering
  \caption{Compression ratios of ZSMILES using cross-dictionaries.}
  \label{tab:zsmilesRatioDataset}
  \begin{tabular}{|c|c|c|c|c|}
    \hline
    \diagbox[width=8em]{\textbf{Train}}{\textbf{Test}} & \textbf{GDB-17} & \textbf{MEDIATE} & \textbf{EXSCALATE} & \textbf{MIXED} \\
    \hline
    \textbf{GDB-17}                                    & 0.33            & 0.60             & 0.60               & 0.55           \\
    \hline
    \textbf{MEDIATE}                                   & 0.46            & 0.29             & 0.29               & 0.35           \\
    \hline
    \textbf{EXSCALATE}                                 & 0.52            & 0.36             & 0.31               & 0.38           \\
    \hline
    \textbf{MIXED}                                     & 0.39            & 0.33             & 0.30               & 0.29           \\
    \hline
  \end{tabular}
\end{table}
In \prettyref{tab:zsmilesRatioDataset} are the compression ratios of ZSMILES when the dictionary is trained on the dataset in the first column and tested on the dataset in the first row.
When the dictionary is trained on GDB-17, MEDIATE, or EXSCALATE, the average compression ratios obtained by compressing other datasets are $0.52$, $0.34$, and $0.39$, respectively.
The compression ratio can vary depending on the chosen training dataset.
The dictionary generated on GDB-17 performs poorly on other datasets, which indicates that GDB-17 consists of homogenous SMILES. 
As expected, using a MIXED dataset gave an average compression ratio of $0.32$, better than all the others.
For this reason, we chose the MIXED dataset for the following experiments.

\textbf{Tools Comparison.}
Finally, we compare the performance of ZSMILES with other state-of-the-art solutions: Bzip2 and FSST, to highlight the benefits of ZSMILES's preprocessing optimizations.
We compare these tools by measuring the compression ratio achieved.
We use this as a test dataset, the MIXED one.
We compressed the same MIXED dataset with each tool and trained ZSMILES's dictionary on the same dataset: we made this decision because FSST constructs a static dictionary for each input, thus allowing us to compare the two approaches fairly.
We expected the binary compression tool to yield the best compression ratio.

\begin{figure}[t]
  \centering
  \includegraphics[width=\columnwidth]{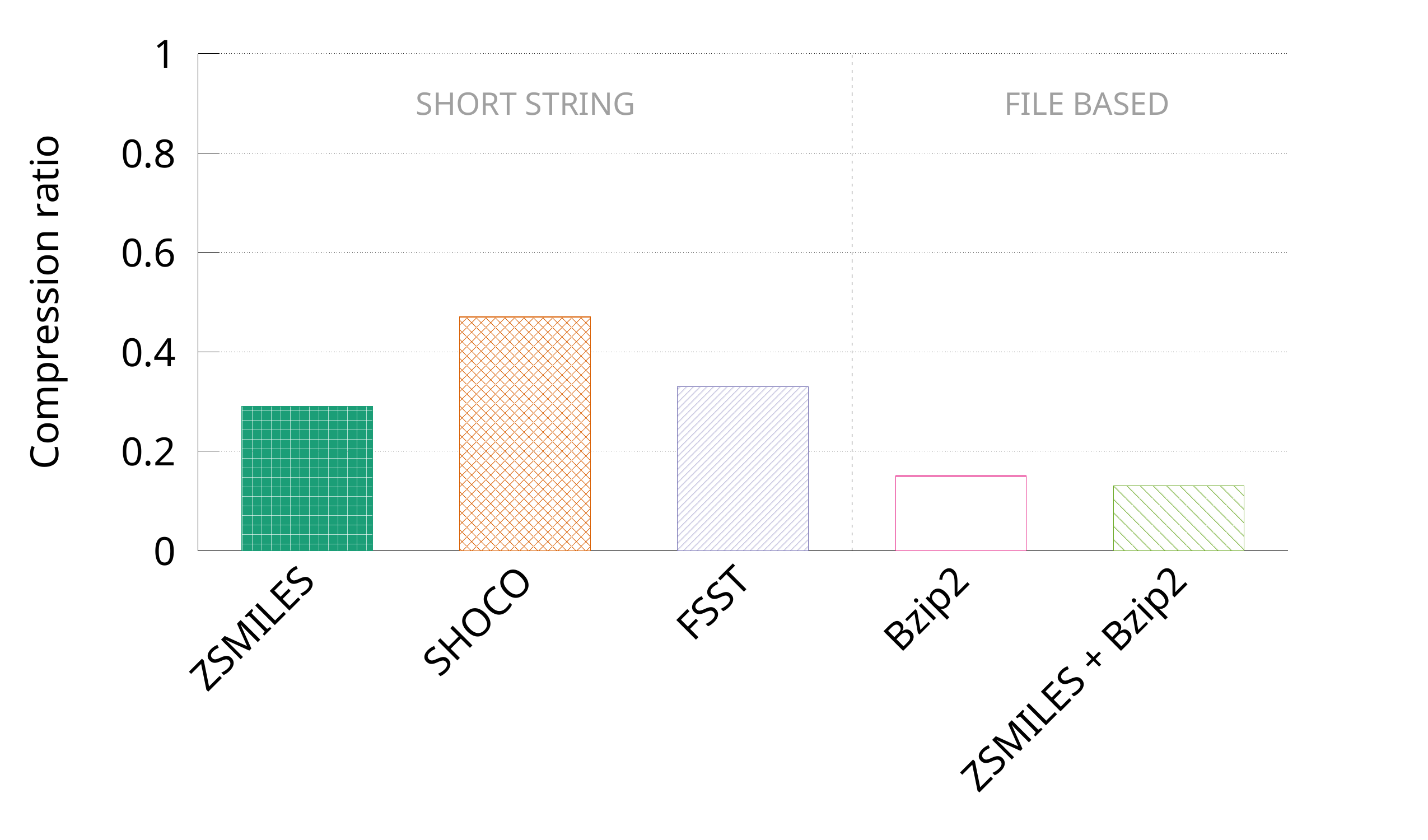}
  \caption{Compression ratios of different tools on a mixed dataset, comparing both short-string and file-based methods.}
  \label{fig:toolsRatio}
\end{figure}

\prettyref{fig:toolsRatio} compares the compression ratio of ZSMILES with FSST, SHOCO, and Bzip2 on a MIXED dataset.
The proposed domain-specific optimizations have been employed only by ZSMILES in this comparison.
On the x-axis is the tool's name used, while on the y-axis is the achieved compression ratio.
\prettyref{fig:toolsRatio} indicates that ZSMILES can provide a good compression ratio while still producing a readable output.
As expected, Bzip2 is the one that performs better, but it does not allow random access or reading of the output since it is binary and the compression is stateful.
ZSMILES performs better than FSST when compressing on the same dataset used for dictionary training.
Using Bzip2 on ZSMILES's output can save even more space as it further compresses the data. It demonstrates the benefits of the preprocessing data step in ZSMILES for the BZIP2 compression algorithm.

\subsection{Performance}
\label{sec:perf}
In this section, we analyze ZSMILES compression and decompression performance: we report ZSMILES execution time by varying the maximum pattern length $L_{max}$ in compression and decompression and with different implementations (C++ and CUDA).
C++ refers to the serial implementation targeting CPUs, while CUDA refers to the parallel one targeting NVIDIA's GPUs.
All execution times reported here consider the execution time of the entire ZSMILES application, and they are normalized to the execution time of the C++ one with the maximum value of $L_{max}$.
We expect the CUDA version to be faster than the C++ one.
We evaluate the execution time of ZSMILES C++ and CUDA version on the Mixed dataset, with different values of $L_{max}$: $5$, $8$, $15$.

\begin{figure}[t]
  \centering
  \begin{subfigure}[b]{1.0\columnwidth}
    \centering
    \includegraphics[width=\textwidth]{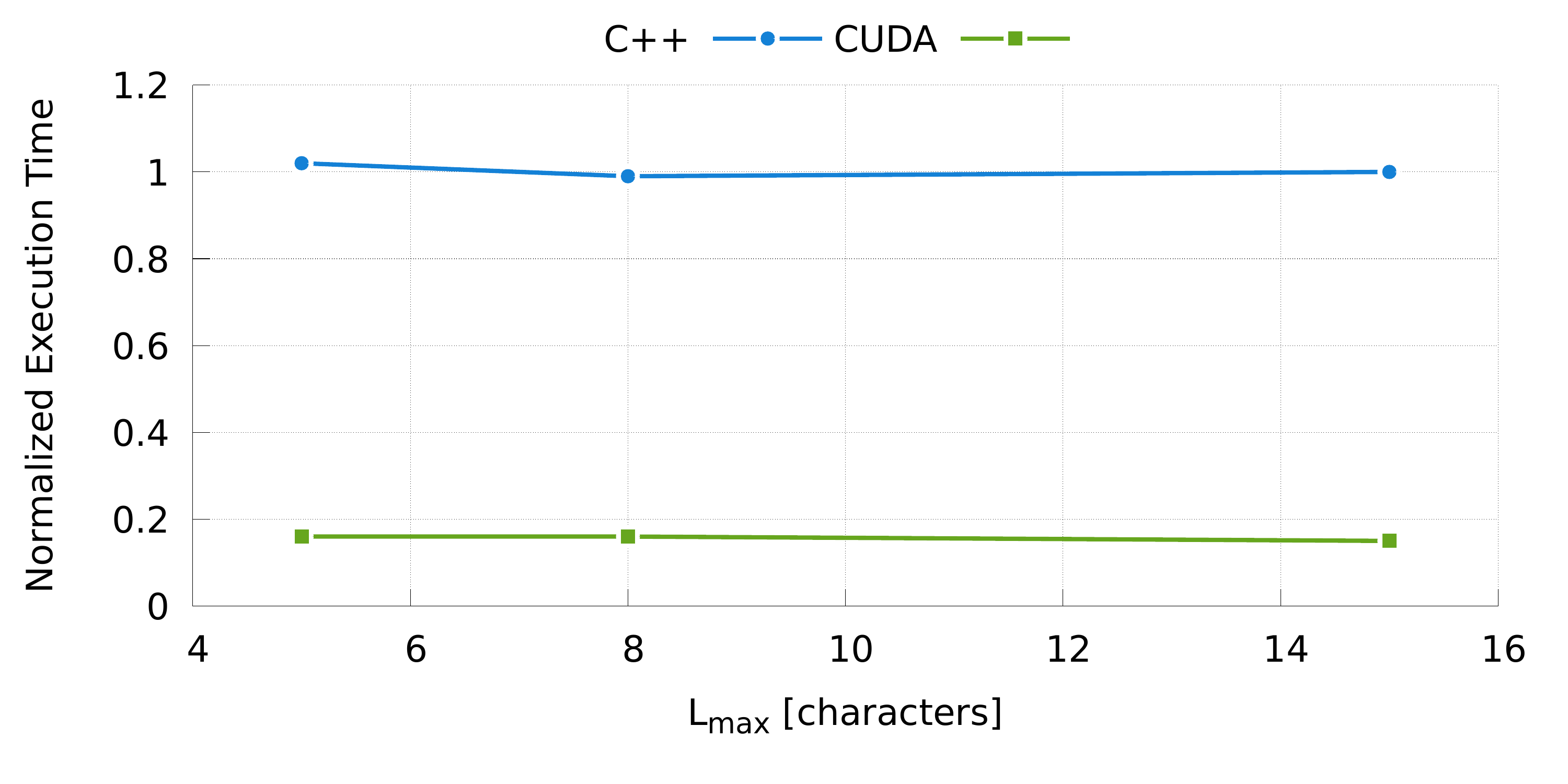}
    \caption{Compression performance.}
    \label{fig:compPerf}
  \end{subfigure}
  \hfill
  \begin{subfigure}[b]{1.0\columnwidth}
    \centering
    \includegraphics[width=\textwidth]{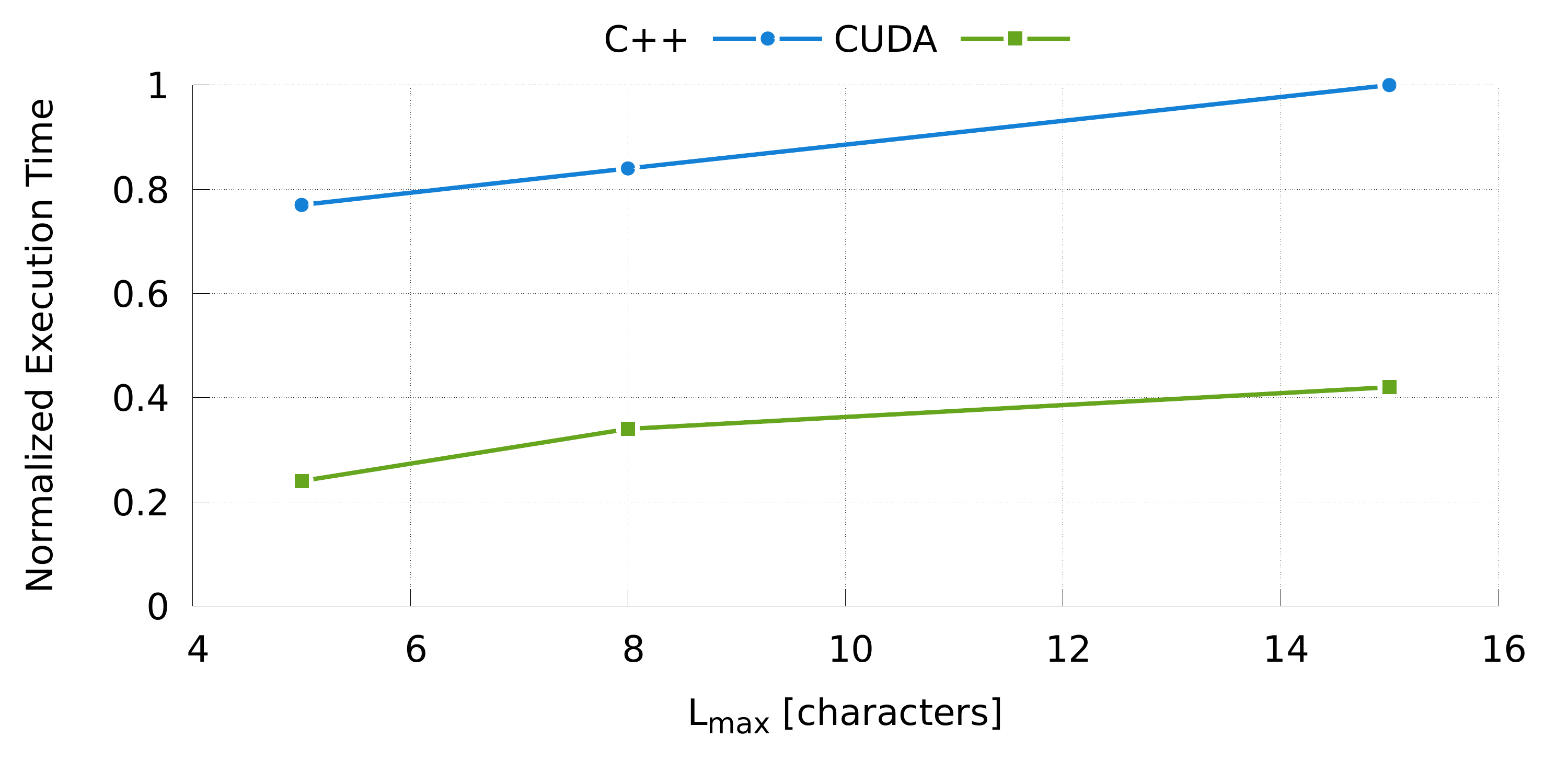}
    \caption{Decompression performance.}
    \label{fig:decompPerf}
  \end{subfigure}
  \caption{ZSMILES normalized execution times of the C++ and CUDA implementation with different $L_{max}$ values.}
  \label{fig:perf}
\end{figure}

\prettyref{fig:perf} reports ZSMILES execution times. $L_{max}$ values are on the x-axis, while ZSMILES normalized execution times are on the y-axis.
\prettyref{fig:compPerf} shows ZSMILES's normalized execution times in compression, while \prettyref{fig:compPerf} in decompression.
As expected, the CUDA implementation is faster in compression and decompression than the C++ one.
In particular, the CUDA version is only $7\times$ faster in compression and $2\times$ in decompression.
We have investigated these speedups further and discovered that for these simple kernels, the bottlenecks are the read-and-write operations on storage: ZSMILES is memory-bound.
Thus, additional C++ or CUDA optimizations have a reduced impact on performance.

\section{Conclusions}\label{sec:conclusions}
Virtual screening campaign requires the storage of large datasets of ligands.
They involve trillions of ligands, which have a big storage footprint.
Some examples in the literature \cite{exascalateBigrun} report the burden of storing large chemical spaces of molecules for extreme-scale campaigns.
In this paper, we have proposed a methodology to store SMILES efficiently, called ZSMILES, by creating a shared dictionary through heuristics, data pre-processing, and dictionary pre-population.
ZSMILES uses an efficient dictionary-based compression algorithm, leveraging SMILES domain knowledge for optimizations and design. 
Given the use case, ZSMILES allows random access to the data and provides a readable output (ASCII format).

ZSMILES can reduce the space required for large datasets by providing a compression ratio of up to $0.29$, showing an improvement over state of the art approach of $\times1.13$.

We propose a C++ and CUDA implementation of ZSMILES, where the CUDA parallel one achieves a speedup of $7\times$ in compression and $2\times$ in decompression compared to the C++ serial implementation.

From experimental results, we have found ZSMILES compression and decompression overhead to be negligible since kernels are inherently memory-bound.



%
\IEEEpeerreviewmaketitle

\ifCLASSOPTIONcompsoc
  \section*{Acknowledgments}
\else
  \section*{Acknowledgment}
\fi
This project has received funding from EuroHPC-JU - the European High-Performance Computing Joint Undertaking - under grant agreement No 956137 (LIGATE). The JU receives support from the European Union's Horizon 2020 research and innovation program and Italy, Sweden, Austria, Czech Republic, and Switzerland.



\bibliographystyle{IEEEtran}
%


\bibliography{main.bib}

\end{document}